\documentclass[apj]{emulateapj}
\usepackage{graphicx}
\usepackage{amsmath}

\newcommand\kms{\ifmmode{\rm km\thinspace s^{-1}}\else km\thinspace s$^{-1}$\fi}
\newcommand\ms{\ifmmode{\rm m\thinspace s^{-1}}\else m\thinspace s$^{-1}$\fi}
\newcommand\msun{\ifmmode{M_{\odot}}\else $M_{\odot}$\fi}
\newcommand\rsun{\ifmmode{R_{\odot}}\else $R_{\odot}$\fi}
\newcommand\mjup{\ifmmode{M_{\rm Jup}}\else $M_{\rm Jup}$\fi}
\newcommand\rjup{\ifmmode{R_{\rm Jup}}\else $R_{\rm Jup}$\fi}
\newcommand{\mysim}{\mathord{\sim}}
\newcommand{\dg}{\ensuremath{^\circ}}
\newcommand\ions[2]{#1$\;${\scshape{#2}}}

\newcommand{\reffigl}[1]{Figure~\ref{fig:#1}}
\newcommand{\refsecl}[1]{\mbox{Section \ref{sec:#1}}}
\newcommand{\refsecls}[2]{\mbox{Sections \ref{sec:#1} and \ref{sec:#2}}}
\newcommand{\refseclr}[2]{\mbox{Sections \ref{sec:#1}$-$\ref{sec:#2}}}

\newcommand{\reftabl}[1]{Table~\ref{tab:#1}}

\submitted{Accepted to ApJ March 30, 2014}

\shorttitle{A Hot Jupiter in the Hyades}
\shortauthors{Quinn et al.}

\begin{document}

\title{HD 285507\lowercase{b}: An Eccentric Hot Jupiter in the Hyades Open Cluster}

\author{
  Samuel N. Quinn\altaffilmark{1,4}, 
  Russel J. White\altaffilmark{1}, 
  David W. Latham\altaffilmark{2},
  Lars A. Buchhave\altaffilmark{2,3},
  Guillermo Torres\altaffilmark{2},
  Robert P. Stefanik\altaffilmark{2},  
  Perry Berlind\altaffilmark{2},
  Allyson Bieryla\altaffilmark{2},
  Michael C. Calkins\altaffilmark{2},
  Gilbert A. Esquerdo\altaffilmark{2},
  Gabor F\H{u}r\'esz\altaffilmark{2},
  John C. Geary\altaffilmark{2},
  and Andrew H. Szentgyorgyi\altaffilmark{2}
}
\altaffiltext{1}{Department of Physics \& Astronomy, Georgia State
  University, 25 Park Place NE Suite 605, Atlanta, GA 30303, USA}
\altaffiltext{2}{Harvard-Smithsonian Center for Astrophysics, 60
  Garden Street, Cambridge, MA 02138, USA}
\altaffiltext{3}{Centre for Star and Planet Formation, Natural History
Museum of Denmark, University of Copenhagen, DK-1350 Copenhagen,
Denmark}
\altaffiltext{4}{NSF Graduate Research Fellow}
\email{quinn@astro.gsu.edu}

\begin{abstract}

We report the discovery of the first hot Jupiter in the Hyades open
cluster. HD 285507b orbits a $V=10.47$ K4.5V dwarf ($M_* =
0.734~\msun$; $R_* = 0.656~\rsun$) in a slightly eccentric
($e=0.086^{+0.018}_{-0.019}$) orbit with a period of
$6.0881^{+0.0019}_{-0.0018}$~days. The induced stellar radial velocity
corresponds to a minimum companion mass of $M_{\rm P} \sin{i} = 0.917
\pm 0.033~\mjup$. Line bisector spans and stellar activity measures
show no correlation with orbital phase, and the radial velocity
amplitude is independent of wavelength, supporting the conclusion that
the variations are caused by a planetary companion. Follow-up
photometry indicates with high confidence that the planet does not
transit. HD 285507b joins a small but growing list of planets in open
clusters, and its existence lends support to a planet formation
scenario in which a high stellar space density does not inhibit giant
planet formation and migration. We calculate the circularization
timescale for HD 285507b to be larger than the age of the Hyades,
which may indicate that this planet's non-zero eccentricity is the
result of migration via interactions with a third body. We also
demonstrate a significant difference between the eccentricity
distributions of hot Jupiters that have had time to tidally
circularize and those that have not, which we interpret as evidence
against Type II migration in the final stages of hot Jupiter
formation. Finally, the dependence of the circularization timescale on
the planetary tidal quality factor, $Q_{\rm P}$, allows us to
constrain the average value for hot Jupiters to be $\log{Q_{\rm P}} =
6.14^{+0.41}_{-0.25}$.

\end{abstract}


\section{Introduction}

The efforts of more than two decades of exoplanet searches have
produced an incredible diversity of discoveries, many of which bear
little similarity to planets in the solar system. These discoveries
provide valuable constraints for our theories of planet formation,
which have evolved to explain the existence of the wide array of
planetary systems observed. One such grouping of planets is the hot
Jupiters, gas giant planets in short period orbits (defined herein as
$M_{\rm P} > 0.3~\mjup$ and $P<10$~days), which occur at rates of
$\mysim 1.2\%$~around Sun-like field stars
\citep[see][]{wright:2012,mayor:2011}. However, more than $60$~years
after their proposed existence \citep{struve:1952} and nearly
$20$~years after the first detection \citep{mayor:1995}, we still lack
a complete description of the hot Jupiter formation process. It is
believed that they form beyond the snow line \citep[located at $\mysim
2.7$~AU for the current-day Sun; e.g.,][]{martin:2013} where there is
a greater reservoir of solids with which to build a massive core
\citep[e.g.,][]{kennedy:2008} before undergoing an inward migration
process, but many questions remain to be answered.

While there are many mechanisms that could cause a gas giant planet to
lose angular momentum and migrate inward \citep[see a discussion
in][]{ford:2008}, two leading ideas are dynamical interactions with a
circumstellar disk \citep[``Type II''
migration;][]{goldreich:1980,lin:1986} and multi-body gravitational
interactions with other planetary \citep[``planet-planet scattering'';
e.g.,][]{rasio:1996,juric:2008} or stellar \citep[Kozai cycles;
e.g.,][]{fabrycky:2007} companions. Migration through a disk must
occur while the gas disk is present \citep[within $\mysim
10$~Myr;][]{haisch:2001} and is expected to preserve near-circular
orbits, but if multi-body interactions are the source of most hot
Jupiters, inward migration may take significantly longer and many of
these planets should initially possess high eccentricity. For
simplicity, we adopt the language of \citet{socrates:2012} and refer
to these multi-body processes as ``high eccentricity migration''
(HEM). Given the different timescales predicted, one direct way to
distinguish between mechanisms would be to search for hot Jupiters
orbiting very young stars ($\lesssim 10$~Myr), and at least one
promising candidate exists orbiting a T-Tauri star \citep[PTFO
8-8695;][]{vaneyken:2012,barnes:2013}. However, the enhanced activity
associated with young stars presents significant challenges for such
surveys \citep[e.g.,][]{bailey:2012}. Alternatively, the dynamical
imprint of HEM on the orbital eccentricity could provide a more
accessible means to observationally constrain migration
\citep[e.g.,][]{dawson:2013,dong:2014}. Subsequent tidal
circularization of the orbits erases this evidence of multi-body
interaction over time, so identifying ``dynamically young'' systems,
for which the system age ($t_{\rm age}$) is less than the
circularization timescale ($\tau_{\rm cir}$), is necessary for such an
investigation. Longer period planets (large $\tau_{\rm cir}$) could
satisfy this requirement, but hot Jupiters with periods of only a few
days are both more common and easier to detect. Young planets (small
$t_{\rm age}$) offer another solution, but field stars tend to be old
and their ages are difficult to estimate accurately
\citep{mamajek:2008}. The ages of open clusters, on the other hand ---
e.g., the $625$~Myr Hyades \citep{perryman:1998} or the $125$~Myr
Pleiades \citep{stauffer:1998} --- are typically precisely known and
can be comparable to (or less than) the circularization timescales of
many hot Jupiters. Consequently, planets in clusters could provide an
avenue to directly measure this observational signature of HEM,
allowing us to determine which process is most important for hot
Jupiter migration.

With the recent discovery of two hot Jupiters \citep{quinn:2012} in
the Praesepe open cluster ($\mysim 600$~Myr), it is now clear that
giant planets can form and migrate in a dense cluster environment.
This had been an open question, as previous exoplanet surveys
targeting open clusters --- including high resolution spectroscopy of
$94$ dwarfs in the Hyades \citep[][hereafter P04]{paulson:2004} and
$58$ dwarfs in M67 \citep{pasquini:2012}, as well as transit
photometry of several other clusters
\citep[e.g.,][]{hartman:2009,pepper:2008,mochejska:2006} --- had not
discovered any planets. However, while small planets do appear to be a
common by-product of star formation, {\em giant} planets are
comparatively rare \citep[e.g.,][]{fressin:2013}. Most cluster surveys
have not been sensitive to the more common smaller planets, so it is
likely that small sample sizes are to blame for the null results
\citep[see also][]{vansaders:2011}. Indeed, two mini-Neptunes recently
discovered \citep{meibom:2013} in NGC 6811 \citep[$\mysim
1$~Gyr,][]{meibom:2011} by the {\em Kepler} spacecraft would not have
been detected by previous surveys, and the results suggest consistency
between occurrence rates in open clusters and the field for planets of
all sizes.

While open cluster planet surveys have recently enjoyed successes in
measuring planet formation rates, they have not yet provided a
convincing constraint for the hot Jupiter migration process, despite
their potential to do so in the long term. Unfortunately, the hot
Jupiters discovered in Praesepe are too old to address timescale
differences between migration mechanisms, and they are not dynamically
young, either. That is, they are close enough to their stars to have
already undergone tidal circularization ($t_{\rm age} > \tau_{\rm
cir}$). To see the dynamical imprint of migration it will thus be
necessary to identify younger hot Jupiters or hot Jupiters on wider
orbits that would still bear the signature of dynamical scattering in
order to help distinguish between migration mechanisms. In hopes of
accomplishing the latter, here we extend our radial velocity (RV)
survey to include more stars with ages and properties comparable to
those in Praesepe.

We describe our sample selection in \refsecl{sample} and outline our
spectroscopic and photometric observations and analysis in
\refsecls{spec}{phot}. We then present evidence for high eccentricity
migration and a constraint on $Q_{\rm P}$ in \refseclr{tcir1}{qp}, and
in \refsecl{disc} we discuss our results.

\section{Sample Selection}
\label{sec:sample}

Hyades cluster members represent an ideal sample for extending our
survey of Sun-like stars in Praesepe \citep{quinn:2012}. The two
clusters are nearly the same age \citep[$625$~Myr and $578$~Myr,
respectively;][]{perryman:1998,delorme:2011}, and both are metal-rich
--- ${\rm [Fe/H]}=+0.13 \pm 0.01$ \citep[Hyades,][]{paulson:2003} and
$+0.19 \pm 0.04$ \citep[Praesepe,][]{quinn:2012}. While P04 found no
planets among $94$ Hyades dwarfs ($74$~FGK stars), a primary
motivation of their survey was to explore the relationship between
planet occurrence and stellar mass, so a number of FGK stars were
omitted in favor of including $20$~M stars. As a result, some suitable
stars have not yet been observed with precise radial velocities.

Our potential targets come from a list assembled by Robert Stefanik
for the CfA Hyades binary survey (R. Stefanik 2013, private
communication). Since $1979$, the CfA has been monitoring over $600$
stars in the Hyades field, extending to a magnitude of
$V=15$. Originally the program consisted of stars drawn from the
Hyades lists of \citet{vanbueren:1952}, \citet{vanaltena:1966}, and
\citet{pels:1975}. Over the years additional stars were added to the
observing program if there was some suggestion that the stars were
possible members based on photometric, proper motion, or radial
velocity investigations of the cluster. Also added to the CfA program
were the companion stars of Hyades visual binaries. From this parent
list, the sample surveyed here was determined after excluding close
($<1\farcs0$) visual binaries revealed by high-resolution imaging
\citep{patience:1998}, the $94$ stars previously surveyed by P04,
rapid rotators ($v\sin(i)>30~\kms$), faint targets ($V>12$), and
spectroscopic binaries (with orbits or long-term trends) from the
literature and the CfA survey. The final target list contained $27$
FGK Hyades members (see \reftabl{stars}).

\begin{deluxetable*}{llrrrrrr}
  \tabletypesize{\small}
  \tablewidth{6.5in}
  \tablecaption{Target List and Observations Summary
    \label{tab:stars}
  }
  \tablehead{
    \colhead{Star} &
    \colhead{Other Name} &
    \colhead{$\alpha$} &
    \colhead{$\delta$} &
    \colhead{$V$} &
    \colhead{$N$} &
    \colhead{$\sigma_{\rm obs}$} &
    \colhead{$v \sin{i}$} \\
    \colhead{} &
    \colhead{} &
    \colhead{(J2000)} &
    \colhead{(J2000)} &
    \colhead{(mag)} &
    \colhead{} &
    \colhead{(\ms)} &
    \colhead{(\kms)}
  }
  \startdata
  J202A    & HIP 15300  & $03:17:22.6$ & $+26:18:55$ & $11.10$ & $ 6$ & $ 39.8$ & $ 2.2$\\
  H24098C  & HD 24098C  & $03:50:13.1$ & $-01:31:08$ & $11.20$ & $ 9$ & $ 40.3$ & $ 0.9$\\
  L7       & HD 283044  & $03:52:40.8$ & $+25:48:16$ & $11.10$ & $ 5$ & $ 12.9$ & $ 2.0$\\
  G7-73C   & \nodata    & $04:01:12.8$ & $+12:05:48$ & $11.30$ & $ 7$ & $ 40.0$ & $ 2.9$\\
  L15      & HD 285507  & $04:07:01.0$ & $+15:20:07$ & $10.47$ & $32$ & $ 89.5$ & $ 3.2$\\
  L18      & HD 284155  & $04:08:36.0$ & $+23:46:07$ & $ 9.42$ & $ 7$ & $ 25.8$ & $ 2.0$\\
  VB13     & HD 26345   & $04:10:42.3$ & $+18:25:24$ & $ 6.59$ & $ 5$ & $ 78.8$ & $28.7$\\
  VB14\tablenotemark{a}							 							      
           & HD 26462   & $04:11:20.2$ & $+05:31:23$ & $ 5.70$ & $24$ & $608.4$ & $17.6$\\
  H111     & HD 286589  & $04:14:51.7$ & $+13:03:18$ & $10.68$ & $ 6$ & $ 22.5$ & $ 2.1$\\
  L26      & HD 285630  & $04:17:25.0$ & $+19:01:47$ & $10.80$ & $ 5$ & $ 23.7$ & $ 2.3$\\
  H210     & HD 286693  & $04:19:36.9$ & $+12:37:27$ & $ 9.80$ & $ 6$ & $  7.5$ & $ 0.6$\\
  H246     & HD 27561   & $04:21:34.7$ & $+14:24:35$ & $ 6.58$ & $ 5$ & $ 30.6$ & $22.1$\\
  L38      & \nodata    & $04:24:07.2$ & $+22:07:08$ & $10.95$ & $ 5$ & $  8.5$ & $ 3.0$\\
  H342     & HD 285742  & $04:25:00.1$ & $+16:59:05$ & $10.37$ & $ 8$ & $ 14.4$ & $ 1.5$\\
  L58      & HD 284455  & $04:26:47.3$ & $+21:14:05$ & $11.24$ & $10$ & $ 21.8$ & $ 2.6$\\
  G7-227   & HD 285765  & $04:27:56.7$ & $+19:03:39$ & $11.36$ & $ 6$ & $ 17.9$ & $ 1.7$\\
  H422     & HD 285804  & $04:28:10.6$ & $+16:28:14$ & $10.72$ & $ 5$ & $ 23.3$ & $ 2.5$\\
  H469     & HD 28406   & $04:29:30.0$ & $+17:51:48$ & $ 6.89$ & $ 5$ & $ 54.3$ & $33.4$\\
  H472     & HD 285805  & $04:29:30.9$ & $+16:14:42$ & $10.63$ & $ 8$ & $ 19.0$ & $ 1.9$\\
  VB86     & HD 28608   & $04:30:56.7$ & $+10:45:07$ & $ 7.03$ & $ 5$ & $ 55.1$ & $30.0$\\
  G8-64    & HD 284785  & $04:47:08.5$ & $+20:52:58$ & $ 9.77$ & $ 6$ & $ 28.8$ & $ 4.3$\\
  AK2-1315 & \nodata    & $04:47:18.0$ & $+06:27:12$ & $11.35$ & $ 7$ & $  7.0$ & $ 2.0$\\
  VB116    & HD 30505   & $04:49:03.0$ & $+18:38:30$ & $ 8.97$ & $ 8$ & $ 28.6$ & $ 4.0$\\
  L96      & HD 286085  & $04:50:00.4$ & $+16:24:45$ & $10.73$ & $ 7$ & $ 28.8$ & $ 3.6$\\
  +23755   & HD 284855  & $04:53:00.6$ & $+23:29:18$ & $10.61$ & $ 7$ & $ 15.2$ & $ 1.7$\\
  L101     & BD +13 741 & $04:57:00.4$ & $+13:54:46$ & $10.86$ & $ 7$ & $ 45.0$ & $ 3.9$\\
  VB128    & HD 31845   & $04:59:44.0$ & $+15:55:03$ & $ 6.75$ & $ 5$ & $ 47.9$ & $34.9$\\
  [-2.4ex]
  \enddata
  \tablenotetext{a}{\footnotesize VB14 is likely a triple star system
    consisting of a single-lined binary and a more distant, yet
    unresolved, companion (see text).}
\end{deluxetable*}

\section{High Resolution Spectroscopy}
\label{sec:spec}
\subsection{Spectroscopic Observations}
\label{sec:obs}

The Tillinghast Reflector Echelle Spectrograph
\citep[TRES;][]{furesz:2008} mounted on the $1.5$~m Tillinghast
Reflector at the Fred L. Whipple Observatory (FLWO) on Mt. Hopkins, AZ
was used to obtain high resolution spectra of Hyades members between
UT 2012 September 23 and 2013 April 08. TRES is a
temperature-controlled, fiber-fed instrument with a resolving power of
$R\mysim 44,000$ and a wavelength coverage of $\mysim
3850$--$9100$~\AA, spanning $51$ echelle orders.

In order to achieve a sensitivity to planets similar to that achieved
for Praesepe stars, we strove to observe each star on two to three
consecutive nights, followed by another two to three consecutive
nights $\mysim 1$~week later \citep[see][]{quinn:2012}. This strategy,
given the RV precision of TRES, is sensitive to planets with masses
greater than about $0.5~\mjup$ and periods up to $10$~days. Though we
were sometimes forced to deviate from the planned observing cadence
because of weather and instrument availability, we were able to obtain
at least $5$ spectra of each of our $27$ stars. Exposure times ranged
from $1$--$25$~minutes, yielding a typical S/N per resolution element
of at least $40$. We also obtained nightly observations of the nearby
IAU RV standard stars HD 38230 ($28\dg$~away) and HD 3765
($54\dg$~away) to help track instrument stability and correct for any
RV zero point drift. Precise wavelength calibration was established by
obtaining ThAr emission-line spectra before and after each spectrum,
through the same fiber as the science exposures.

Compared to the sample in P04, we typically have fewer observations
per star. However, we were able to observe at higher cadence because
of the flexible TRES queue schedule. They used a more sensitive
instrument on a bigger telescope \citep[Keck-HIRES;][]{vogt:1994}, but
we were able to overcome much of the aperture difference because the
stars are bright. Moreover, the RV precision of P04 was mainly limited
by the intrinsic stellar jitter, not the instrumental precision or
photon statistics. Our sample does include some stars with more rapid
rotation (up to $\mysim 30~\kms$), so to make a fair comparison we
ignore the stars with companions or with $v\sin{i}>20~\kms$, and find
our average measured velocity dispersion to be $\mysim 23~\ms$, which
is comparable to that of P04 ($\mysim 16~\ms$). The details of our
spectroscopic analysis are described below.

\subsection{Spectroscopic Reduction and Cross-correlation}
\label{sec:reduction}

Spectra were optimally extracted, rectified to intensity versus
wavelength, and for each star the individual spectra were
cross-correlated, order by order, using the strongest exposure of that
star as a template \citep[for details, see][]{buchhave:2010}. We
typically used $\mysim 25$~orders ($\mysim4200$--$6500$~\AA),
rejecting those plagued by telluric absorption, fringing far to the
red, and low S/N far to the blue. For each epoch, the
cross-correlation functions (CCFs) from all orders were added and fit
with a Gaussian to determine the relative RV for that epoch. Using the
summed CCF rather than the mean of RVs from each order places more
weight on orders with high correlation coefficients.  Internal error
estimates (which include, but may not be limited to, photon noise) for
each observation were calculated as $\sigma_{\rm int}=$~RMS({\bf {\em
v}})$/\sqrt{N}$, where {\bf {\em v}} is the RV of each order, $N$ is
the number of orders, and RMS denotes the root-mean-squared velocity
difference from the mean.

To evaluate the significance of any potential velocity variation, we
compared the observed velocity dispersions ($\sigma_{\rm obs}$) to the
combined measurement uncertainties, which we assumed stem from three
sources: (1) internal error, $\sigma_{\rm int}$ (described above), (2)
night-to-night instrumental error, $\sigma_{\rm TRES}$, and (3) RV
jitter induced by stellar activity, $\sigma_*$.

Before assessing the instrumental error, we used observations of HD
38230 and HD 3765 to correct for any systematic velocity shifts
between runs, calculated in the following way. First, the median RV of
each of the two standard stars was calculated for each run, resulting
in two sets of run-to-run offsets. We took the error-weighted mean of
these offsets to be the final run-to-run offsets, which were then
applied to our Hyades data. We note that the instrument has been
remarkably stable during the span of our observations, with run-to-run
offsets similar to their uncertainties, typically less than
$3~\ms$. After run-to-run corrections, the RMS of the standard star
RVs was $10.5~\ms$ with internal errors of only $7.6~\ms$. Since we
expect negligible stellar jitter for the RV standards, the
instrumental floor error should be given by $\sigma_{\rm
  TRES}=\sqrt{\sigma_{\rm obs}^2 - \sigma_{\rm int}^2}$. Thus, we
adopted a night-to-night instrumental error of $\sigma_{\rm
  TRES}=7.2~\ms$. In order to reduce identification of false signals
caused by noisy stars, we set $\sigma_* = 16~\ms$ \citep[the average
  velocity RMS for all Hyads surveyed by][]{paulson:2004} in our
initial analysis of the $27$ Hyades stars.

\begin{deluxetable}{rrr}[!t]
  \tabletypesize{\small}
  \tablewidth{2.75in}
  \tablecaption{Relative Radial Velocities of HD 285507
    \label{tab:rvs}
  }
  \tablehead{
    \colhead{BJD} &
    \colhead{RV} &
    \colhead{$\sigma_{\rm RV}$\tablenotemark{a}} \\
    \colhead{($-2,456,000$)} &
    \colhead{(\ms)} &
    \colhead{(\ms)}}
  \startdata
  $ 196.96928$ & $  42.3$ & $ 10.0$ \\
  $ 207.93330$ & $ 180.1$ & $  9.4$ \\
  $ 208.90481$ & $  61.6$ & $ 10.7$ \\
  $ 209.86006$ & $   7.7$ & $ 10.7$ \\
  $ 210.82123$ & $  92.9$ & $  9.3$ \\
  $ 211.89901$ & $ 252.5$ & $ 15.0$ \\
  $ 223.88429$ & $ 198.3$ & $  7.0$ \\
  $ 224.83584$ & $ 266.6$ & $ 10.4$ \\
  $ 225.82361$ & $ 228.3$ & $  8.0$ \\
  $ 226.83416$ & $ 113.0$ & $ 10.6$ \\
  $ 227.79041$ & $  16.9$ & $  7.6$ \\
  $ 228.80807$ & $  65.0$ & $  9.6$ \\
  $ 229.78857$ & $ 179.7$ & $  9.3$ \\
  $ 234.94762$ & $  72.6$ & $  9.3$ \\
  $ 235.89156$ & $ 167.1$ & $  8.0$ \\
  $ 236.82198$ & $ 256.2$ & $  8.8$ \\
  $ 237.85251$ & $ 226.8$ & $  9.2$ \\
  $ 238.92242$ & $ 106.0$ & $  8.0$ \\
  $ 263.79884$ & $  85.0$ & $  8.6$ \\
  $ 264.81227$ & $  12.2$ & $  8.9$ \\
  $ 267.84157$ & $ 257.2$ & $  8.5$ \\
  $ 268.94476$ & $ 169.6$ & $  9.9$ \\
  $ 282.78804$ & $   1.3$ & $  8.5$ \\
  $ 283.74106$ & $  72.7$ & $ 11.1$ \\
  $ 293.85762$ & $ 100.8$ & $  7.2$ \\
  $ 325.71304$ & $   0.0$ & $  7.0$ \\
  $ 350.64048$ & $  96.2$ & $ 10.0$ \\
  $ 351.70178$ & $ 209.5$ & $ 12.7$ \\
  $ 358.63303$ & $ 249.5$ & $  8.8$ \\
  $ 362.68212$ & $  68.4$ & $ 16.8$ \\
  $ 374.66505$ & $   7.5$ & $  9.5$ \\
  $ 390.62275$ & $ 216.4$ & $ 12.6$ \\
  [-2.4ex]
  \enddata
  \tablenotetext{a}{\footnotesize The errors listed here are internal
    error estimates, but in the orbital solution we include an
    instrumental floor error of $7.2~\ms$, added in quadrature with
    the internal errors.}
\end{deluxetable}

Accounting for internal errors, instrumental jitter, and stellar
noise, we constructed a $\chi^2$ fit of each star's RVs assuming a
constant velocity, and then calculated $P(\chi^2)$, the probability
that the observed $\chi^2$ value would arise from a star of constant
RV. Given constraints imposed by telescope time, our threshold for
further follow-up was $P(\chi^2)<0.001$ (i.e., $99.9\%$ confidence of
variability). Two stars met this criterion. The first, HD 26462,
initially showed variation suggestive of a planetary or brown dwarf
companion ($\mysim 1~\kms$), but subsequent observations revealed a
larger variation and a strong correlation between the line broadening
and the radial velocities. We concluded that two sets of spectral
lines were present and that the true variation is much larger than a
few $\kms$, but diluted by the blended set of lines. HD 26462 is most
likely a hierarchical triple system composed of a single-lined binary
and a more distant stellar companion, and we will discuss it in more
detail in a subsequent paper about the stellar populations of Praesepe
and the Hyades. The second star to meet our variability threshold was
HD 285507, which also stood out obviously by eye as having significant
RV variations after just $3$ observations. We continued to monitor it
over the rest of the season, obtaining $32$ epochs spanning $194$
days. The radial velocities are presented in \reftabl{rvs}, and we
discuss the system in detail in the following sections. None of the
other $25$ stars meet our criterion, and the distribution of their
$P(\chi^2)$ values is roughly uniform, as one would expect for a
sample of constant stars with appropriate error estimates.

\begin{deluxetable}{lcc}
  \tabletypesize{\small}
  \tablewidth{3in}
  \tablecaption{Stellar and Planetary Properties
    \label{tab:props}
  }
  \tablehead{
    \colhead{Orbital parameters} &
    \colhead{}
  }
  \startdata
  $P$ (days)                                 & $\hspace{19mm}6.0881 \pm 0.0018$     \\
  $T_{\rm c}$ (BJD)                          & $\hspace{10mm}2456263.121 \pm 0.029$ \\
  $K$ (\ms)                                  & $\hspace{16.2mm}125.8 \pm 2.3$       \\
  $e$\tablenotemark{a}                       & $\hspace{19mm}0.086 \pm 0.019$       \\
  $\omega$ (deg)\tablenotemark{a}            & $\hspace{17.5mm}182 \pm 11$          \\
  $\gamma_{\rm rel}$ (\ms)                   & $\hspace{16.2mm}143.9 \pm 1.6$       \\
  $\gamma_{\rm abs}$ (\kms)\tablenotemark{b} & $\hspace{17.7mm}38.149 \pm 0.023$    \\
  \hline
  \vspace{0.1mm} \\
  [-1.5ex]
  \multicolumn{2}{l}{Physical properties} \\
  \vspace{0.1mm} \\
  [-2.1ex]
  \hline
  \vspace{0.1mm} \\
  [-1.5ex]
  $M_*$ (\msun)\tablenotemark{c}             & $\hspace{19mm}0.734 \pm 0.034$         \\
  $R_*$ (\rsun)\tablenotemark{c}             & $\hspace{19mm}0.656 \pm 0.054$         \\
  $T_{\rm eff,*}$ (K)\tablenotemark{c}       & $\hspace{16mm}4503^{+85}_{-61}$        \\
  $\log{g}_*$ (dex)\tablenotemark{c}         & $\hspace{18.5mm}4.670^{+0.051}_{-0.058}$ \\
  $v\sin{i}$ (\kms)                          & $\hspace{19mm}3.2 \pm 0.5$             \\
  $[Fe/H]$ (dex)\tablenotemark{c}            & $\hspace{16.7mm}+0.13 \pm 0.01$          \\
  $Age$ (Myr)\tablenotemark{c}               & $\hspace{17.5mm}625 \pm 50$              \\
  $M_{\rm P} \sin{i}$ (\mjup)                & $\hspace{19mm}0.917 \pm 0.033$         \\
  [-2.4ex]
  \enddata
  \tablenotetext{a}{\footnotesize
    The MCMC jump parameters included the orthogonal quantities
    $\sqrt{e} \cos{\omega}$ and $\sqrt{e} \sin{\omega}$, but we report
    the more familiar orbital elements $e$ and $\omega$.
  }
  \tablenotetext{b}{\footnotesize
    The absolute center-of-mass velocity has been shifted to the RV
    scale of \citet{nidever:2002}, on which the velocities of HD 3765
    and HD 38230 are $-63.202~\kms$~and $-29.177~\kms$, respectively.
  }
  \tablenotetext{c}{\footnotesize
    From the final isochrone fits (\refsecl{props}). [Fe/H] and age
    were fixed to values determined for the cluster
    \citep{paulson:2003,perryman:1998}.
  }
\end{deluxetable}

\subsection{Orbital Solution}
\label{sec:orbits}

We used a Markov Chain Monte Carlo (MCMC) analysis to fit Keplerian
orbits to the radial velocity data of HD 285507, fitting for orbital
period $P$, time of conjunction $T_{\rm c}$, the radial velocity
semi-amplitude $K$, the center-of-mass velocity $\gamma_{\rm rel}$,
and the orthogonal quantities $\sqrt{e}\cos{\omega}$ and
$\sqrt{e}\sin{\omega}$, where $e$ is eccentricity and $\omega$ is the
argument of periastron. We adopted errors corresponding to the extent
of the central $68.3\%$ interval of the MCMC posterior distributions.

\begin{figure}[!tb]
\epsscale{1.2}
\plotone{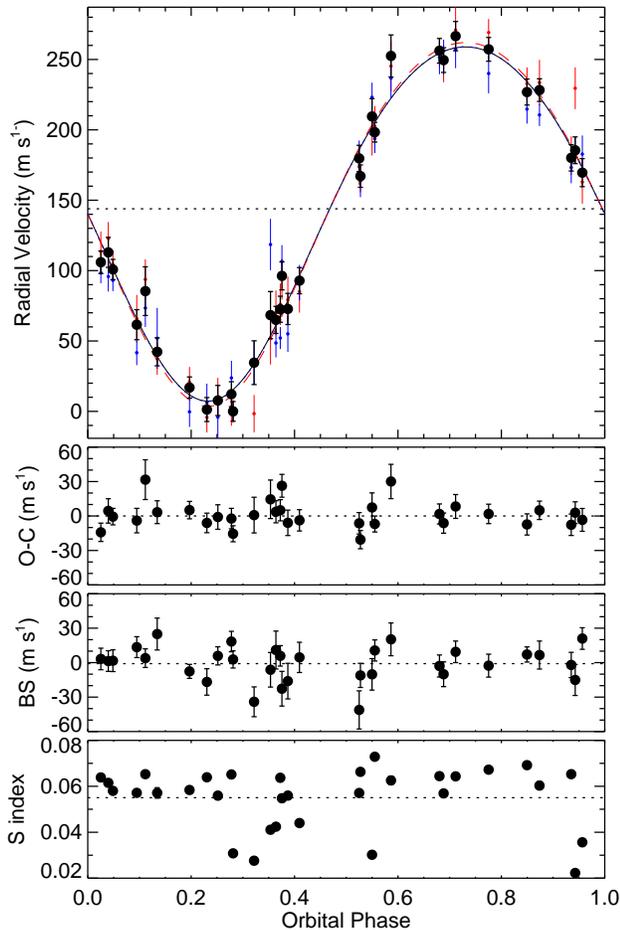}
\caption{Orbital solution for HD 285507b. The panels, from top to
bottom, show the relative RVs, best-fit residuals, bisector span
variations, and relative S index values. In the top panel the large
black points are the final RVs, but also plotted are the RVs derived
from the blue and red orders of the spectrum, showing agreement at
different wavelengths (see \refsecl{bisectors}). RV error bars
represent the internal errors, and do not include astrophysical or
instrumental jitter, although $7.2~\ms$ instrumental jitter was added
to the orbital fit. The solid black curve shows the best-fit orbital
solution (and the blue and red dashed curves show the fits to the blue
and red RVs). The blue curve is nearly indistinguishable from the
black curve. The orbital parameters are listed in \reftabl{props}.
\label{fig:orbit}}
\end{figure}

The RV errors did not require the addition of stellar jitter in order
to obtain a good fit ($\chi^2_{\rm red}=1$), so we set $\sigma_*=0$ in the
orbital solution. We report the best fit orbital parameters in
\reftabl{props} and plot the best fit orbit in \reffigl{orbit}.

Because a modest non-zero eccentricity causes only a small deviation
from a circular orbit, we also investigated whether there exists
correlated RV noise (e.g., due to surface activity and rotation) on
timescales similar to the orbital period using the method of
\citet{winn:2010}. Such noise could in principle cause small
deviations from a circular orbit that might be interpreted as orbital
eccentricity. To rule out this scenario, we fit a circular orbit and
performed the test on the residuals to that solution. We found no
evidence for correlated noise on any timescale.

\subsection{Tests for a False Positive}
\label{sec:bisectors}

HD 285507 is slowly rotating \citep[$P_{\rm
rot}=11.98$~days;][]{delorme:2011}, no X-ray emission was detected by
{\em ROSAT} \citep{stern:1995}, and no stellar jitter term was
required to obtain a good fit to the radial velocities, all of which
are suggestive of a chromospherically inactive star. Nevertheless, to
rule out false positive scenarios in which the observed RV variations
are caused by stellar activity or stellar companions, we used our
observations of HD 285507 to search for spectroscopic signatures that
correlate with the orbital period.

If the RV variations were caused by a background blend
\citep{mandushev:2005} or star spots \citep{queloz:2001}, we would
expect the shape of the star's line bisector to vary in phase with the
radial velocities. A standard prescription for characterizing the
shape of a line bisector is to measure the relative velocity at its
top and bottom; this difference is referred to as a line bisector span
\citep[see, e.g.,][]{torres:2005}. To test against background blends
or star spots, we computed the line bisector spans for all
observations of HD 285507. As illustrated in \reffigl{orbit}, the
bisector span variations are small ($\sigma_{\rm BS} = 15~\ms$) and
they are not correlated with the observed RV variations, having a
Pearson {\em r} value of only $0.06$.

\begin{figure}[!tb]
\epsscale{1.2}
\plotone{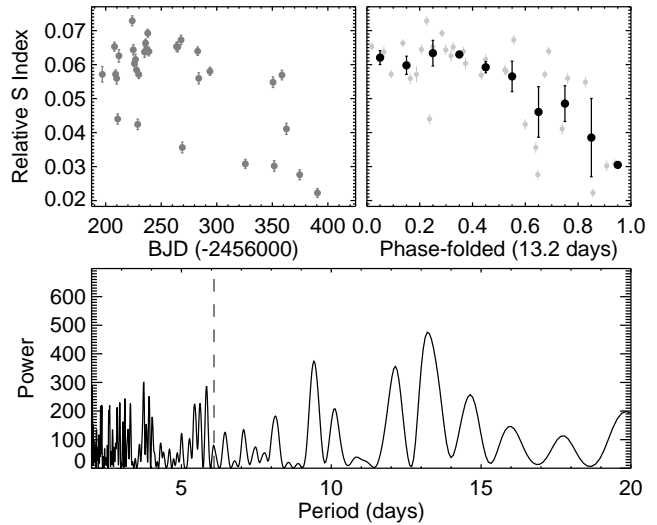}
\caption{ 
Top left: the stellar activity, as characterized by the relative S
index measured from our spectra of HD 285507. Bottom: the
Lomb--Scargle periodogram indicates that there may be significant
periodicity on timescales similar to the stellar rotation period of
$11.98$~days \citep{delorme:2011}, but not at the RV period of
$6.09$~days (dashed line). Top right: the data have been folded onto
the highest peak, $13.2$~days, and binned for illustration.
\label{fig:sind}}
\end{figure}

For each spectrum we also computed the S index --- an indicator of
chromospheric activity in the \ions{Ca}{ii} H\&K lines. We follow the
procedure of \citet{vaughan:1978}, but we note that our S indices are
not calibrated to their scale; these are relative
measurements. Correlation between S index and orbital phase might be
expected if the apparent RV variations were activity-induced, but as
shown in \reffigl{orbit}, there is no such correlation (Pearson
$r=0.17$). Instead, there may be significant periodicity in the S
indices at $12$~or $13$~days (\reffigl{sind}), which is similar to the
published rotation period of $11.98$~days. Our data set is too sparse
to claim a detection of the rotation period from the activity
measurement, but we can see there is no power at the observed orbital
period of $6.088$~days.

Finally, if spots were the source of the variation, we might also
expect the RV amplitude to be wavelength dependent because contrast
between the spot and the stellar photosphere is wavelength
dependent. We derived RVs for the blue and red orders separately
(weighted mean wavelengths of $\lambda_{\rm blue} = 4967$~\AA~and
$\lambda_{\rm red} = 5845$~\AA), and find the amplitudes to be
consistent at the level of $0.5\sigma$ ($3~\ms$,
\reffigl{orbit}). This agreement between the red and blue amplitudes
is encouraging, but is not conclusive by itself. We generally expect
large amplitude differences between the optical and infrared for
spot-induced RVs because the spot contrast can change drastically over
that wavelength range. The expected amplitude difference for our
smaller ($\mysim 1000$~\AA) wavelength span, on the other hand, is
more uncertain because the local wavelength dependence of the RV
amplitudes is itself dependent upon the (unknown) temperature
difference between spot and photosphere
\citep[e.g.,][]{reiners:2010,barnes:2011}. The simulations of
\citet{reiners:2010} indicate that the amplitude difference might be
detectable if the spot contrast is low, but not if it is high. Even
this is not certain, though, as other authors \citep{desort:2007}
predict a $10\%$ drop in amplitude between blue and red for high
contrast spots on solar-type stars. Regardless, from these results and
the work of \citet{saar:1997}, we estimate that to induce the observed
RV amplitude ($125~\ms$) given $v \sin{i} \approx 3.2~\kms$, the spot
would have to cover $\mysim 20\%$ of the visible stellar surface for
low contrast spots ($\Delta_{T_{\rm eff}} \approx 200$~K) or $\mysim
5\%$ for high contrast spots ($\Delta_{T_{\rm eff}} \approx
1500$~K). Large and high contrast spots are more likely to appear on
very magnetically active stars \citep[e.g.,][]{bouvier:1995}, and we
have no evidence for strong magnetic activity. Furthermore, for such
spot configurations, the RV-bisector correlation should be strong
\cite[e.g.,][]{mahmud:2011}, and we observe no correlation.

We conclude from the evidence presented above that the observed RV
variation is not caused by spots, but is the result of an orbiting
planetary companion.

\subsection{Stellar and Planetary Properties}
\label{sec:props}

We used the spectroscopic classification technique Stellar Parameter
Classification \citep[SPC;][]{buchhave:2012} to determine the
effective temperature $T_{\rm eff}$, surface gravity $\log{g}$,
projected rotational velocity $v\sin{i}$, and metallicity [m/H] of HD
285507. In essence, SPC cross-correlates an observed spectrum against
a grid of synthetic spectra, and uses the correlation peak heights to
fit a $3$-dimensional surface in order to find the best combination of
atmospheric parameters ($v\sin{i}$ is fit iteratively since it is only
weakly correlated to changes in the other parameters). We used the CfA
library of synthetic spectra, which are based on Kurucz model
atmospheres \citep{kurucz:1992} calculated by John Laird for a
linelist compiled by Jon Morse. Like other spectroscopic
classification techniques, SPC can be limited by degeneracy between
parameters, notably $T_{\rm eff}$, $\log{g}$, and [m/H], but in this
case we can enforce the known cluster metallicity \citep[$+0.13 \pm
0.01$,][]{paulson:2003} to partially break that degeneracy.

To determine the physical stellar parameters, we utilized the
Dartmouth \citep{dotter:2008}, Yonsei-Yale \citep{yi:2001}, and Padova
\citep{girardi:2000} stellar models. Applying an observational
constraint on the size of the star --- imposed indirectly by the
spectroscopic $T_{\rm eff}$, $V$ magnitude \citep[$V=10.473 \pm
0.012$,][]{roser:2011}, and distance \citep[$41.34 \pm
3.61$~pc;][]{vanleeuwen:2007} --- and enforcing the age
\citep[$625$~Myr;][]{perryman:1998} and metallicity of the Hyades, we
determined the best fit mass and radius for each of the three
isochrones. All three results agreed to within $3\%$ in mass and $5\%$
in radius, and although the resulting $\log{g}$ values indicated by
the isochrones were consistent with the spectroscopically determined
value, the temperatures were nominally discrepant at the $2\sigma$
level. It is possible that, for stars of this mass and age, the
stellar models and/or SPC suffer from a systematic bias not reflected
in the formal errors. Given that the exact stellar parameters have
little bearing on the results presented in this paper, we choose to
simply caution the reader and inflate the errors on stellar mass and
radius by a factor of two. We adopted the mean mass and radius from
the three isochrone fits ($M_* = 0.734 \pm 0.034~\msun$, $R_* = 0.656
\pm 0.054~\rsun$), where the uncertainties listed here are the
inflated statistical errors. \reftabl{props} lists all of the stellar
and planetary properties. We note that our adopted temperature
($4503$~K) is consistent with previous estimates of the spectral type
\citep[e.g., K5,][]{nesterov:1995}, and using the spectral
type/temperature relations assembled in \citet{kraus:2007}, we
estimate a more precise spectral type of HD 285507 to be K4.5.

\begin{figure}[!tb]
\epsscale{1.2}
\plotone{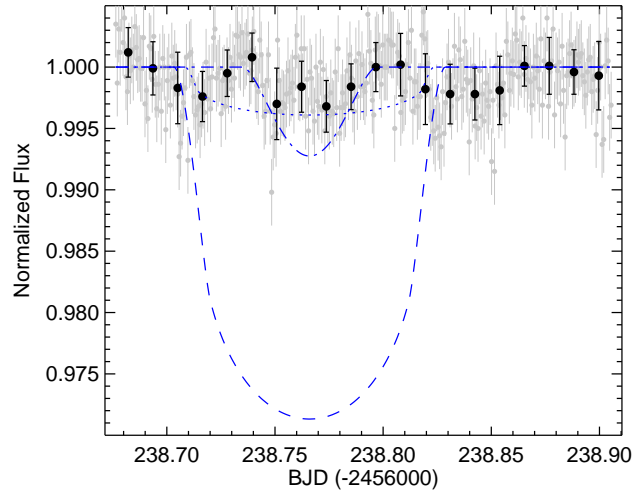}
\caption{
Light curve of HD 285507, showing the individual data (small gray
circles), the binned data (large black circles), and three simulated
transit models at the predicted time of transit (all of which are
rejected as bad fits to the data): a central transit of a $0.95~\rjup$
planet (dashed line), a central transit of a $0.35~\rjup$ planet
(dotted line), and a grazing transit ($b=1$) of a $0.95~\rjup$ planet
(dash-dotted line). The latter two cases mark simulated detection
limits for the observed data quality (see \refsecl{phot}). The
uncertainty in the time of conjunction at this epoch is only about
$1$~hr ($\mysim0.04$~days).
\label{fig:lc}}
\end{figure}

\section{Stellar Inclination and a Search for Transits}
\label{sec:phot}

Since the rotation period of HD 285507 is $11.98$~days and we have
estimates for $R_*$ and $v\sin{i}$, we can in principle calculate the
inclination of the stellar spin axis. In practice, the fractional
uncertainty on $v\sin{i}$ is large (not because the absolute
uncertainty is large, but because the value is small), and the
inclination can only be constrained to be $i > 72\dg$. This does not
exclude an edge-on stellar equator ($i=90\dg$), and because hot
Jupiters orbiting cool stars ($\lesssim 6250$~K) tend to be
well-aligned with the stellar spin axis \citep[see,
e.g.,][]{albrecht:2012}, an inclination of $\mysim 90\dg$ would make a
transit more likely a priori. However, even an inclined stellar spin
axis would not preclude transits of HD 285507, as there is evidence to
suggest that young planets tend to be more misaligned than old planets
\citep[e.g.,][]{triaud:2011}. In addition to providing a radius
measurement for HD 285507b, transits of this relatively young planet
orbiting a cool star could be valuable to the interpretation of these
intriguing correlations.

With this in mind, we conducted photometric monitoring of HD 285507
with KeplerCam on the FLWO $1.2$~m telescope at the predicted time of
conjunction on UT 2012 November 7. KeplerCam is a monolithic,
4k$\times$4k Fairchild 486 chip with a $23' \times 23'$ FOV and a
resolution of $0.''336~{\rm pixel}^{-1}$. We used a Sloan $i$ filter
with exposure times of $45$~s and readout time of $12$~s,
obtaining a total of $334$ images over $5.5$~hours. We reduced the raw
images using the IRAF package \verb+mscred+, and performed aperture
photometry with SExtractor \citep{bertin:1996}.

The resulting light curve showed no sign of a transit, but to
determine our detection sensitivity, we simulated transits --- using
the routines of \citet{mandel:2002} and a quadratic limb darkening law
from \citet{claret:2012} --- and injected them into our observed
data. For each injected transit, we compared the mean flux in transit
(for all points between mid-ingress and mid-egress) to the mean flux
out of transit. If the two differed by more than $1\sigma$, we
classified that transit as detected. From this, we can rule out a
central transit of objects larger than $0.35~\rjup$ (see
\reffigl{lc}). If HD 285507b were to transit, then the derived minimum
mass would be the true planetary mass. Under this assumption and using
the mass-radius-flux relation from \citet{weiss:2013}, we would then
expect its radius to be $\mysim 0.95~\rjup$, much larger than our
sensitivity limit. We can also rule out all transits of a $0.95~\rjup$
planet with impact parameters $b < 1$ (i.e., all but the most extreme
grazing transits). Using the final ephemeris, the transit center
should have occurred $2.22$~hr after our observations began
($T_{\rm c}=2456238.7686$~BJD), with an uncertainty of $\mysim
1$~hr, so it is unlikely that a transit occurred outside our
observing window.

\section{Evidence for Dynamical Scattering of HD 285507\lowercase{b}}
\label{sec:tcir1}

The discovery of a hot Jupiter in the Hyades open cluster brings the
total number of short period giant planets in clusters to $3$. Of
these, however, HD 285507b is unique in that it is the only one that
is definitively eccentric; the two planets in Praesepe are consistent
with having circular orbits. As described in the introduction, the
non-zero eccentricity of HD 285507b could be a tracer of its migration
history if the planet is dynamically young (i.e., $\tau_{\rm cir} >
t_{\rm age}=625$~Myr). If it is dynamically old, then the orbit should
have already circularized, and establishing a credible link between
the eccentricity and the migration process becomes more difficult. In
planet--planet scattering for example, if the outer planet gets
ejected during the scattering event as is expected, then one must
invoke a separate mechanism to excite eccentricity again after
circularization. To put it differently, if HD 285507b is dynamically
young, then planet--planet scattering is sufficient (but not
necessary) to explain the observations; if the planet is dynamically
old, planet--planet scattering is neither sufficient nor necessary. To
test these scenarios, we estimate $\tau_{\rm cir}$ using the equation
given by \citet{adams:2006}:

\begin{align}
\tau_{\rm cir} = &~1.6~{\rm Gyr} \times \left(\frac{Q_{\rm P}}{10^6}\right) \times \left(\frac{M_{\rm P}}{M_{\rm Jup}}\right) \times \left(\frac{M_*}{\msun}\right)^{-1.5} \nonumber \\
                 & \times \left(\frac{R_{\rm P}}{R_{\rm Jup}}\right)^{-5} \times \left(\frac{a}{0.05~{\rm AU}}\right)^{6.5}
\end{align}

\noindent where $Q_{\rm P}$ is the planetary tidal quality factor (a
measure of the efficiency of tidal dissipation within the
planet). Note that $\tau_{\rm cir}$ scales linearly with $Q_{\rm P}$,
which is unknown to within an order of magnitude. The Jupiter--Io
interaction does provide the constraint $6 \times 10^4 < Q_{\rm Jup} <
2 \times 10^6$ \citep{yoder:1981}, but $Q_{\rm P}$ is likely dependent
upon temperature, composition, rotation, and internal structure, all
of which may be quite different for hot Jupiters. $Q_{\rm
P}\approx10^6$, which we adopt herein, is a fiducial value often
assumed for short period giant planets \citep[for a more detailed
discussion of tidal dissipation, see, e.g.,][]{ogilvie:2004}.

Since HD 285507b does not transit, we do not know $R_{\rm P}$ and
measure only a minimum mass, $M_{\rm P} \sin{i}$. However, we can
calculate the expectation value of $\sin{i}$ for randomly oriented
orbits to determine the most likely mass, and then use the giant
planet mass--radius--flux relation derived by \citet{weiss:2013} to
estimate the planetary radius:

\begin{equation}
\frac{R_{\rm P}}{R_\oplus} = 2.45~\left(\frac{M_{\rm P}}{M_\oplus}\right)^{-0.039}\left(\frac{F}{{\rm erg~s^{-1}~cm^{-2}}}\right)^{0.094}
\end{equation}

\noindent where $F$ is the time-averaged incident flux on the
planet. Since $R_{\rm P}$ depends only weakly on $M_{\rm P}$, assuming
an inclination is not likely to introduce a large radius error ---
there is only a $1\%$ difference in derived radius between edge-on and
average-inclination orientations.

Under these assumptions, we find $\tau_{\rm cir} \approx 11.8$~Gyr ---
much larger than the age of the cluster. Note that this holds true
even for the full range of $Q_{\rm Jup}$ (corresponding to $700~{\rm
Myr} < \tau_{\rm cir} < 22.6$~Gyr). We conclude that HD 285507b is
dynamically young. While it is tempting to thus proclaim that
migration has occurred via a HEM mechanism, recall that this is not a
necessary condition for a dynamically young planet with an eccentric
orbit. For any individual planet, non-zero eccentricity could also be
the result of continued interaction with an undetected planetary or
stellar companion, a recent close stellar encounter, or even modest
eccentricity excitation via Type II migration
\citep[e.g.,][]{dangelo:2006}. Only analysis of a population of
planets can provide meaningful insight into the migration process in
this manner. Therefore, we turn to the literature for ages and
circularization timescales of the known sample of hot Jupiters.

\section{Evidence for Dynamical Scattering Among Known Exoplanets}
\label{sec:tcir2}
\subsection{Description of the Analysis}

\begin{figure*}[!t]
\epsscale{0.9}
\plotone{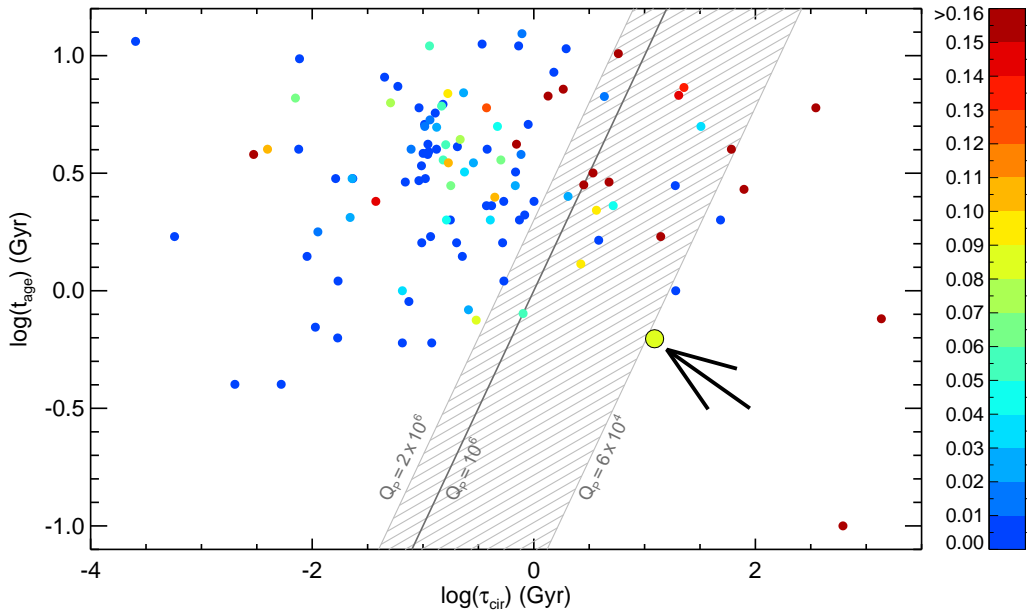}
\caption{
Age vs. circularization timescale for short period ($P<10$~days)
massive ($M>0.3~\mjup$)~planets, assuming a tidal quality factor
$Q_{\rm P} = 10^6$~for all planets. The solid dark line indicates
where $t_{\rm age}=\tau_{\rm cir}$; planets to the left of the line
are expected to have undergone tidal circularization. We also plot a
shaded region to show the estimated uncertainty in this boundary given
the range of $Q_{\rm P}$~values consistent with observations of the
Jupiter--Io interaction \citep[see ][]{yoder:1981}. The data points
are colored according to their eccentricity, and HD 285507b is the
large outlined circle indicated by the arrow. There is a hint that the
right side is populated by preferentially eccentric (red) planets and
the left side by preferentially circular (blue) planets. This is
explored further in \reffigl{edist}.
\label{fig:tcir}}
\end{figure*}

\begin{figure}[!ht]
\epsscale{1.2}
\plotone{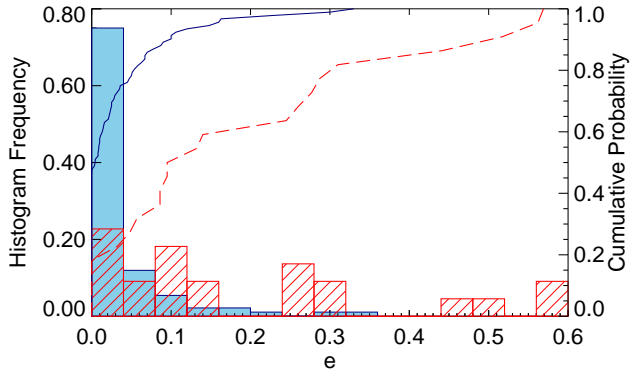}
\caption{
Eccentricity histograms and cumulative distributions of hot Jupiters
that have (filled blue; left of the thick solid line in \reffigl{tcir})
and have not (striped red; right of the line) been tidally
circularized. As in \reffigl{tcir}, we have assumed $Q_{\rm P}=10^6$
for all planets. A KS test rejects the hypothesis that the two
subsamples are drawn from the same distribution with $99.997\%$
confidence.
\label{fig:edist}}
\end{figure}

To search for dynamical imprints of migration among known hot Jupiters
($M_{\rm P} > 0.3~\mjup$, $P < 10$~days), we follow the prescription
described above to calculate $\tau_{\rm cir}$ (and $M_{\rm P}$ and
$R_{\rm P}$ for non-transiting planets). We adopt ages,
eccentricities, planetary masses and radii, and stellar masses, radii,
and temperatures from the literature.\footnote{All values were
obtained from The Extrasolar Planets Encyclopaedia,
www.exoplanet.eu. Only planets with ages listed on exoplanet.eu are
included in this analysis.} In \reffigl{tcir}, we plot $t_{\rm age}$
versus $\tau_{\rm cir}$ for this sample. While the figure is
complicated by the uncertainties already discussed as well as poorly
constrained ages and potential biases in measuring modest
eccentricities
\citep[e.g.,][]{shen:2008,pont:2011,zakamska:2011,wang:2011}, there is
a hint that the points to the right of the circularization boundary
are preferentially eccentric and the ones to the left are
preferentially circular. If HEM were responsible for the final stages
of hot Jupiter migration, this would be expected; planets get
scattered inward on highly eccentric orbits and circularize over
time. If Type II migration were responsible, we should expect very
little difference between the eccentricity distributions to the right
and left of the boundary; ordered migration through a gas disk should
largely preserve circular orbits, so subsequent tidal interactions
would not change the population significantly. In \reffigl{edist}, we
plot the eccentricity histograms and cumulative distributions for the
two populations, which contain $92$ and $22$ planets (as shown in
\reffigl{tcir}). To quantify the difference between them, we ran a
Kolmogorov--Smirnov (KS) test. The KS $p$-value is the likelihood that
the two subsamples came from the same parent distribution, and in this
case, $p=3.3\times 10^{-5}$. We conclude that the two distributions do
not come from the same parent distribution, with $\mysim
99.997\%$~confidence, and infer that high eccentricity migration
mechanisms play a significant role in hot Jupiter migration. We also
ran an Anderson--Darling (AD) test, which is similar to a KS test, but
is more sensitive to differences in the distribution tails. The AD
test indicates an even greater significance that the two distributions
do not come from the same parent distribution of eccentricities.

\subsection{The Effect of Measurement Errors}
\label{sec:errors}

As noted previously, ages and eccentricities can be difficult to
determine for many of these systems, so it is important to consider
what effect uncertainties may have on the significance of our
result. Ages of field stars can be estimated by many techniques,
including gyrochronology, stellar activity, lithium abundance, and
isochrone fitting. However, ages derived from multiple techniques do
not always agree, and when they do agree, the allowed range of ages
can still be quite large. Likely for this reason, The Extrasolar
Planets Encylcopaedia does not report uncertainties on the age (when
age is reported at all). \citet{mamajek:2008} claim a precision of
$\mysim0.2$~dex in their activity--age relation, and while not all
stars in this sample have ages derived in this manner, we believe this
to be an appropriate approximate error for isochrone fitting as well,
which is one of the more widespread techniques employed to determine
ages. We therefore adopt this as a typical error in our analysis.

Eccentricity errors are similarly heterogeneously reported in the
literature, especially for nearly circular orbits. Some authors assume
zero eccentricity in such cases for simplicity, which introduces a
bias toward smaller values, while others report upper limits or a
measured eccentricity. When a small measured value is reported, it may
be biased toward larger eccentricities depending on the details of the
fitting. Rather than worry about potential conflicting biases in a
heterogeneous set of eccentricities and associated errors, we assume a
constant eccentricity error of $0.05$ for all planets in our
sample. We also assume errors of $10\%$ on stellar and planetary
masses and radii, $3\%$ on semi-major axis, and $100$~K on stellar
effective temperature. These values are minor contributors to
uncertainty in the analysis, but they do have a small effect on the
derived circularization timescales.

Using the above errors, we redraw our sample $10^4$ times. For each of
these simulated data sets, we run a KS test to determine the
significance of the difference in populations, resulting in a
distribution of $10^4~p$-values. The median of this distribution is
$2.7 \times 10^{-4}$, or a $99.973\%$ confidence (nearly $4\sigma$)
that the two samples come from different parent
distributions. Furthermore, we find that even for eccentricity errors
as large as $0.1$ (which is unrealistically large for nearly all
planets in the sample), the KS significance remains greater than
$3\sigma$. From this, we conclude that even with conservatively
large errors, our result holds: dynamically young planets have larger
eccentricities, which suggests HEM mechanisms contribute significantly
to hot Jupiter migration.

\section{A Constraint on the Tidal Quality Factor $Q$ for Hot Jupiters}
\label{sec:qp}

Until now we have been assuming $Q_{\rm P} = 10^6$ to determine which
planets are dynamically young (right side of the circularization
boundary in \reffigl{tcir}) and which are dynamically old (left side),
allowing us to draw conclusions about the migration process. If we
instead {\em start} with the assumption that planets migrate inward in
possession of some intial eccentricity (rather than being gently
shepherded on circular orbits through the gas disk), we can invert the
problem to place a constraint on the tidal quality factor $Q_{\rm
P}$. As we vary $Q_{\rm P}$, the circularization boundary changes
location, and the difference between the two populations should be
maximized (and the KS $p$-value minimized) for the correct (average)
hot Jupiter $Q_{\rm P}$. Note that it does not matter what fraction of
hot Jupiters has undergone HEM -- if any fraction has, then the
minimum $p$-value should occur when the circularization boundary is in
the correct place.

\begin{figure}[!ht]
\epsscale{1.2}
\plotone{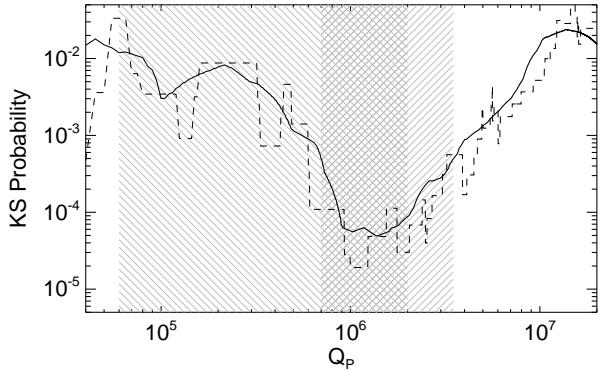}
\caption{
KS $p$-values from comparisons of the eccentricity distributions of
hot Jupiter systems with $t_{\rm age} < \tau_{\rm cir}$ to those with
$t_{\rm age} > \tau_{\rm cir}$, as a function of the assumed $Q_{\rm
P}$ (dashed line); the solid line shows the same result after a boxcar
filtering. Since the minimum $p$-value should occur for the most
realistic $Q_{\rm P}$ (see \refsecl{qp}), we can constrain $Q_{\rm P}$
for a typical hot Jupiter to be between about $7.8 \times 10^5$ and
$3.5 \times 10^6$ (right shaded region encompassing the broad
minimum). The Jupiter--Io constraint (left shaded region) only overlaps
partially with the position of the minimum at $Q_{\rm P}=1.39 \times
10^6$, and given the expectation that $Q$ will be larger for hot
Jupiters, we do not apply this constraint to our adopted range for
$Q_{\rm P}$.
\label{fig:ks}}
\end{figure}

\reffigl{ks} shows the results of this experiment. A value on the
order of $10^6$ is preferred, which seems to rule out much of the
parameter space consistent with Jupiter's tidal quality factor (and
validates our assumption of $Q_{\rm P} = 10^6$). The discrete data
points do not produce a smooth distribution, so finding the minimum is
not straightforward. As such, we smooth it using a moving average with
a boxcar filter of size [$\frac{2}{3}Q_{\rm P},\frac{3}{2}Q_{\rm
    P}$]. Quantitative confidence limits on the minimum are difficult
to assess (the vertical axis is {\em not} a probability associated
directly with $Q_{\rm P}$), but we take approximate upper and lower
limits on $Q_{\rm P}$ to be those for which the smoothed p-value is
$10$ times its smoothed minimum value, which occurs for $Q_{\rm
  P}=1.39 \times 10^6$. For each simulated data set described in
\refsecl{errors}, we calculate $Q_{\rm P}$ this way and find the
median to be similar ($1.12 \times 10^6$), but with smaller
errors. This is not surprising, as the statistical errors from the
simulations are akin to a standard deviation of the mean whereas the
errors derived from a single p-value versus $Q_{\rm P}$ experiment are
akin to a sample standard deviation. To describe the population, we
therefore adopt the latter and suggest an appropriate tidal quality
factor for a typical hot Jupiter is $\log{Q_{\rm P}} =
6.14^{+0.41}_{-0.25}$ (see \reffigl{ks} for a visual representation).

\section{Summary and Conclusions}
\label{sec:disc}

Our discovery of a hot Jupiter in the Hyades bolsters the statistics
of short period giant planets in open clusters (of which three are now
known). Including the \citet{paulson:2004} null result in the Hyades
($0$ planets among $74$ FGK stars), our Hyades sample ($1$ in $26$),
and an updated census of Praesepe including unpublished data from our
second year of observations ($2$ in $60$), a total of $3$ out of $160$
stars host a hot Jupiter. After correcting for completeness and
calculating Poisson errors following the prescription in
\citet{gehrels:1986}, we find a hot Jupiter frequency of
$1.97^{+1.92}_{-1.07}\%$ in the metal-rich Praesepe and Hyades open
clusters. However, giant planet occurrence scales with metallicity
approximately as $10^{2 {\rm [Fe/H]}}$ \citep{fischer:2005}. If we
take ${\rm [Fe/H]} \approx +0.15$~as representative of the combined
Praesepe and Hyades sample, the solar-metallicity-adjusted hot Jupiter
frequency in clusters is $0.99^{+0.96}_{-0.54}\%$. Although more
discoveries are needed to reduce the uncertainty, this is in good
agreement with the frequency for field stars \citep[$1.20 \pm 0.38
\%$;][]{wright:2012}, and improves the evidence that planet frequency
is the same in clusters and the field.

A primary motivation for the search for young planets is that their
ages are comparable to the timescale of migration. Thus, the orbital
properties of such planets may still bear the dynamical signature of
this process. Since different migration mechanisms are predicted to
produce hot Jupiters on different timescales and with different
orbital eccentricities, we can use the properties of young hot
Jupiters (and their existence at various ages) to determine the
process by which they migrate. The ages of the cluster planets
discovered thus far do not place a strong direct constraint on the
{\em timescale} of migration (we know only that the process took less
than $600$~Myr), but the newly discovered planet in the Hyades holds a
clue its dynamical history. HD 285507b has a long circularization
timescale, so its non-zero eccentricity may be a remnant of the
migration process, which would suggest planet-planet scattering or
Kozai cycles have played a role in its orbital evolution. There is no
observational evidence for a third body, but one cannot be excluded
either. The RV timespan is not long enough to rule out a second giant
planet (which also could have been ejected during scattering), and
imaging by \citet{patience:1998} only rules out companions more
massive than $0.13~\msun$ with projected separations $5$--$50$~AU.

Applying this idea more broadly, we have compared ages and
circularization timescales for all known hot Jupiters and find
evidence for two families of planets, distinguished by their orbital
properties: (1) mostly circular orbits for the ``dynamically old''
planets, those with $\tau_{\rm cir} < t_{\rm age}$, and (2) a range of
eccentricities for ``dynamically young'' planets, those with
$\tau_{\rm cir} > t_{\rm age}$. If Type II migration were the leading
driver of hot Jupiter migration, both dynamically young and old
planets should have circular orbits. We thus conclude that HEM is
important for producing hot Jupiters. However, we can only say that
these planets have experienced dynamical stirring {\em at some point},
and do not suggest that this evidence shows Type II migration to be
unimportant. On the contrary, as shown by simulations time and again,
Type II migration is almost certainly important to orbital evolution
before the gas disk dissipates, but we suggest that for a large
fraction of hot Jupiter systems, planet-planet scattering or the Kozai
mechanism is responsible for the final stages of inward migration. A
larger sample of dynamically young (non-circularized) planets may
allow us to determine what that fraction is. Since few hot Jupiters
have circularization timescales greater than $1$~Gyr, a good way to
enhance this sample is to continue finding young planets.

That HEM mechanisms play an important role in hot Jupiter migration
has already been suggested, and is supported by a rich data set of
stellar obliquity measurements in hot Jupiter systems \citep[see][for
a recent discussion]{albrecht:2012}. In addition to the excitation of
orbital eccentricity, dynamical encounters with a third body are
expected to produce a range of orbital inclinations, although tidal
interactions with the host star may realign the systems over
time. These inclinations can be measured precisely, most notably via
the Rossiter--McLaughlin effect \citep{rossiter:1924,mclaughlin:1924},
and the results of such studies parallel those presented in this
paper: systems for which the tidal timescale is short tend to be
well-aligned, and those for which the timescale is long display high
obliquities. \citet{albrecht:2012} do caution that stars and their
disks may be primordially misaligned for reasons unrelated to hot
Jupiters, but we see no obvious reason for this to influence the
eccentricities. Migration through multi-body dynamical interactions,
on the other hand, could explain both the inclined orbits and high
eccentricities observed in systems that have not yet experienced
significant tidal interactions. Whether that process is primarily
planet-planet scattering or the Kozai effect remains to be determined,
and it is likely that more data will be needed to properly answer this
question.

Finally, the tidal circularization boundary that separates the
dynamically young and old populations of hot Jupiters is sensitive to
the choice of the planetary tidal quality factor, $Q_{\rm P}$, so we
have leveraged this dependence to constrain the typical value for hot
Jupiters to be $\log{Q_{\rm P}} = 6.14^{+0.41}_{-0.25}$. $Q_{\rm P}$
has wide-ranging implications, e.g., for simulating orbital evolution
\citep{beauge:2012} or modeling the inflated radii of hot Jupiters
\citep{bodenheimer:2003}, but has thus far proven difficult to
constrain observationally. While our result still includes substantial
uncertainty and will not be applicable to any one planet, it can be
applied to these problems in a statistical sense. Moreover, it offers
a path forward: as our sample of longer period and young hot Jupiters
grows, the determination of $Q_{\rm P}$ using this method should
improve.

\acknowledgements 
We thank Josh Winn, Tsevi Mazeh, and an anonymous referee for
insightful comments and discussion, and Greg Feiden for producing a
Dartmouth isochrone appropriate for our analysis. This research has
made use of The Extrasolar Planets Encyclopaedia. The material herein
is based upon work supported by the National Aeronautics and Space
Administration (NASA) under grant No. NNX11AC32G issued through the
Origins of Solar Systems program. S.N.Q. is supported by an NSF
Graduate Research Fellowship, grant DGE-1051030. D.W.L. acknowledges
partial support from NASA's {\em Kepler} mission under cooperative
agreement NNX11AB99A with the Smithsonian Astrophysical
Observatory. G.T. acknowledges partial support from NSF grant
AST-1007992.

{\it Facilities:} \facility{FLWO:1.5m (TRES)}, \facility{FLWO:1.2m
(KeplerCam)}

\end{document}